# Assessing Electricity Network Capacity Requirements for Industrial Decarbonisation in Great Britain

Ahmed Gailani[*a, b], Peter Taylor [a,b,c]

[a] Energy Transition and Net-Zero Research Group, School of Chemical and Process Engineering, University of Leeds, Leeds, LS2 9JT, United Kingdom

[b] The UK Energy Research Centre, University College London, London, WC1H 0NN

[c] Sustainability Research Institute, School of Earth and Environment, University of Leeds, Leeds, LS2 9JT, United Kingdom

[*]**Email: ahmedgailani90@gmail.com**

# Abstract

Decarbonising the industrial sector is vital to reach net zero targets. The deployment of industrial decarbonisation technologies is expected to increase industrial electricity demand in many countries and this may require upgrades to the existing electricity network or new network investment. While the infrastructure requirements to support the introduction of new fuels and technologies in industry, such as hydrogen and carbon capture, utilisation and storage are often discussed, the need for investment to increase the capacity of the electricity network to meet increasing industrial electricity demands is often overlooked in the literature. This paper addresses this gap by quantifying the requirements for additional electricity network capacity to support the decarbonisation of industrial sectors across Great Britain (GB). The Net Zero Industrial Pathways model is used to predict the future electricity demand from industrial sites to 2050 which is then compared spatially to the available headroom across the distribution network in GB. The results show that network headroom is sufficient to meet extra capacity demands from industrial sites over the period to 2030 in nearly all GB regions and network scenarios. However, as electricity demand rises due to increased electrification across all sectors and industrial decarbonisation accelerates towards 2050, the network will need significant new capacity (71 GW + by 2050) particularly in the central, south, and north-west regions of England, and Wales. Without solving these network constraints, around 65% of industrial sites that are large point sources of emissions would be constrained in terms of electric capacity by 2040. These sites are responsible for 69% of industrial point source emissions.

# 1. Introduction

Industrial commodities such as steel, cement, and chemicals are essential for a vibrant economy. Their demand and production have significantly increased over recent decades resulting in a significant rise in global greenhouse gas (GHG) emissions. In 2023, the industry sector accounted for 38.2% of total global final energy consumption and 26.5% of energy-related $CO_2$ emissions [1]. In the UK, the industrial sector is less important than in some countries, but industrial GHG emissions still account for around 14% of total emissions [2]. As industrial commodities are needed to enable the net zero transition, it is vital to reduce the associated emissions to allow their continued production while meeting climate targets.

The UK government's industrial decarbonisation strategy published in 2021 aimed to reduce sectoral emissions by two-thirds in 2035 and more than 90% by 2050 [3]. Government modelling suggested that this could be achieved by a combination of resource and energy efficiency, significant deployment of low carbon hydrogen and carbon capture, utilisation and storage (CCUS) technologies and, to a lesser extent, electrification of some industrial processes. Delivering this strategy would require significant investment in nationally critical infrastructure for hydrogen production, transportation and storage along with CCUS transportation and storage, as without these "national networks", residual industrial emission in 2050 would more than triple [3].

However, the slow pace in the development and delivery of CCUS and hydrogen projects globally has re-shifted attention in some countries, including the UK, towards the role of electrification to decarbonise industry, particularly as there have been promising developments in electrified technologies to decarbonise energy-intensive industries [4, 5]. Any increased deployment of industrial electrification technologies, as well as plans for other electricity-intensive decarbonisation options such as CCUS and green hydrogen, will likely increase electricity demand. For instance, the International Energy Agency expects the share of electricity in industrial energy demand to double by 2050 compared to 2023 under a net zero scenario [1]. The authors in [6] examined the impacts on EU electricity demand of electrifying key energy intensive industries in Europe and found that the annual electricity demand could increase by 44% (rising by 1200 TWh). This increase in electricity demand requires new investment or upgrade to the electricity network resulting in connection cost implications for industrial sites [7-9].

Increased electricity use by industry is not the only source of additional demands on the network. Many countries in Europe are already starting to see the electrification of heat and transport and therefore the network will need upgrading in a timely manner to support increased electricity flows from a range of end-use sectors. In the UK, there are already reports of industrial sites facing significant delays for network upgrade or connections. A recent survey found that electricity network constraints is a significant risk to industrial decarbonisation with delays of 2-12 years to access the network [10]. Furthermore, a UK government consultation found that getting network access is the second most important barrier to electrification with connection delays of up to 5 years [11]. In response the government, industry and the energy regulator are working together to improve and accelerate grid connections across the network [12].

While previous studies have explored the impact of increased industrial electrification on electricity demand and highlighted the impact that delayed network connections could have on industry decarbonisation plans, there is currently a lack of quantitative analysis for the UK (and other countries) that brings together spatially disaggregated information on the available network headroom in the distribution network with the location of industrial sites. Such analysis could help to inform future UK government policy on electrification as this issue has not so far been sufficiently considered in its own modelling, and industry is concerned that without grid reform the problem could be exacerbated by new policies to promote electrification[11] .  The results can also highlight to governments, and policy and research institutes the benefits of taking a whole system approach when planning for energy infrastructure.

This paper, therefore, fills a gap in our current understanding by exploring the potential impact of electricity network constraints on the decarbonisation of industry sectors in Great Britain (GB)[1]. We explore the regional level of network capacity needed to decarbonise industrial sites, the percentage and type (cluster/dispersed) of industrial sites that are constrained, and proportion of current emissions from these constrained sites and the resulting impact on net zero goals due to network capacity constraints.

---

[1] The analysis is performed for Great Britain (comprising England, Scotland and Wales) and not the United Kingdom as these countries form a single electricity market. The electricity market in Northern Ireland is separate.

## 2. Methods

Figure 1 provides an overview of the methods used in this paper. The current and future electricity demand headroom data is collected for substations across all Distribution Network Operators (DNOs) in GB up to the year 2050. This headroom data takes account of all sources of future electricity demand identified by the DNOs, including that arising from increased penetration of electric vehicles and future deployment of heat pumps. Then, the Net Zero Industrial Pathways (NZIP) model [13] is used to project the future electricity demand for industrial sites as they decarbonise. Both datasets then were made consistent for the purpose of the analysis in this paper, for example, by using unified MW unit, location information, or scenario names. Next, the data on network demand headroom and the additional electrical capacity needed by industrial sites (both point and non-point sources[2]) are aggregated to 11 Regions in GB (9 regions in England[3], plus Wales and Scotland) and then compared to obtain the level of network constraints by region.

Since the location is known for industrial sites that are identified as point-sources, spatial optimisation was used to find the nearest substation to each of these sites by finding the Haversine distance to minimise the cost of network upgrades (cables, etc.) and losses. Then, the substation capacity was optimally allocated by allowing sites with small capacities be accommodated first. The validation of this approach is shown in Figure A 1 (Appendix).

---

[2] Point sources are industrial sites with significant greenhouse gas emissions for which the geographical location is known due to various legal reporting requirements (e.g. the UK Emissions Trading Scheme). Non-point sources are industrial sites with lower emissions and for which the only locational information available is the region in which they are based.

[3] There were nine Government Offices for the Regions (GOR) established across England in 1994 which were abolished in 2011. Due to the requirement to maintain a region-level geography for statistical purposes, the Government Statistical Service Regional and Geography Committee agreed that from 1 April 2011, the former GORs should be simply referred to as 'regions'.

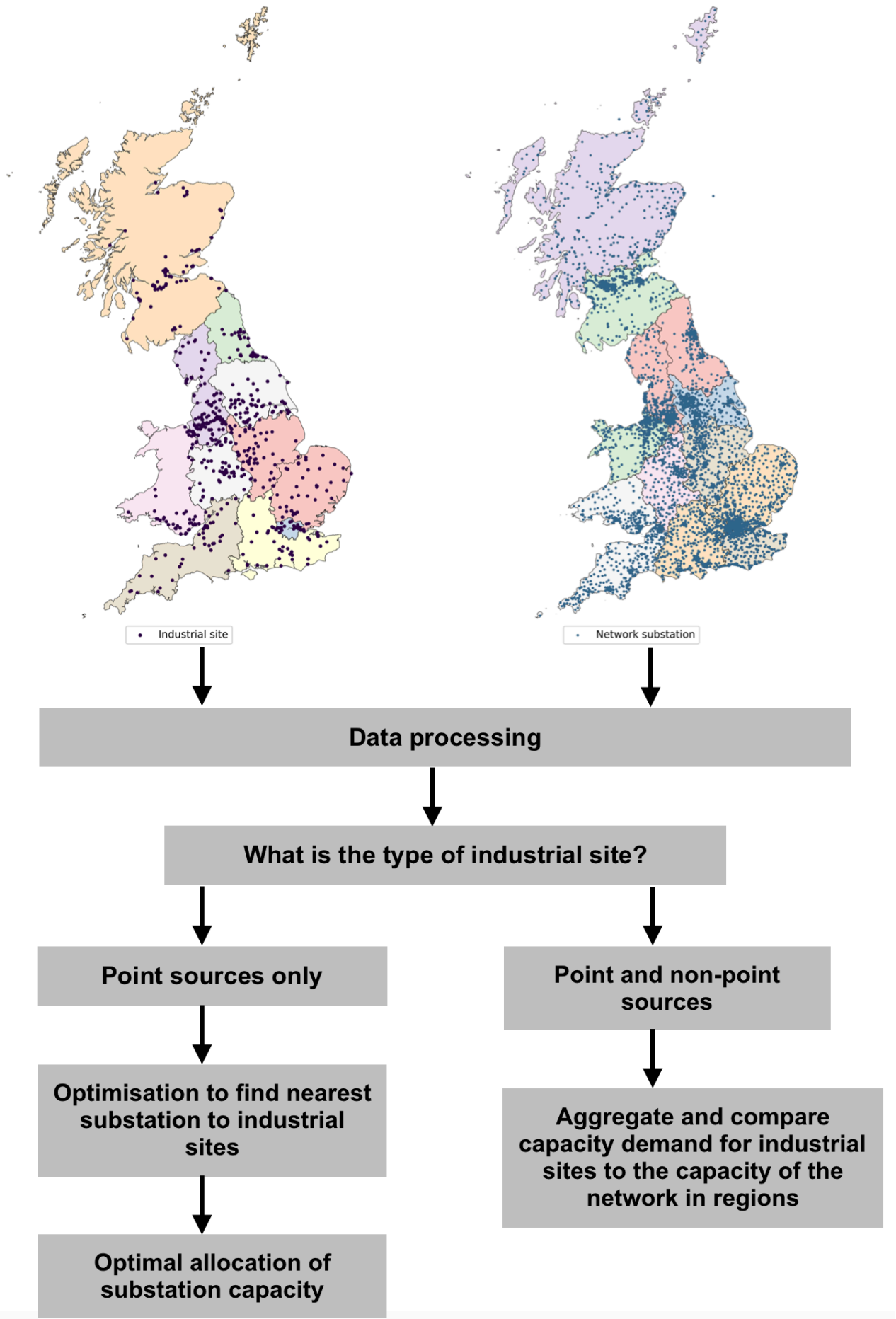

Figure 1 Methods used in this study.

## 2.1. Network headroom data

The electricity demand headroom data were firstly collected for all DNOs in GB [14-19] from their network development plans. The electricity headroom, in general, is the asset rated capacity minus the current or future power flow in that asset. These network development plans are used by the DNOs to identify where upgrades to the distribution system will be needed in the future. As such, they take account of short-term plans to upgrade to the network (typically to around 2030) but not any expansion of network capacity in the longer term. They also consider the availability of headroom in the high-voltage transmission network and how this impacts the headroom available at the distribution level.

They then use a scenario-based approach to consider how electricity demand may increase in the future, particularly as a result of increased deployment of heat pumps and electric vehicles. As such, the network headroom data takes account of the holistic demand assumed by DNOs including from other sectors such as building and transport.

The data represents the thermal demand headroom for all network substations up to 66 kV. As the data comes in different spatial formats, units, temporal coverage, and scenarios, reasonable assumptions were made to make the data as consistent as possible. For example, a power factor of 90% was assumed to convert the apparent power in MVA to active power in MW and in cases where DNOs have 2031 as a data point rather than 2030, then 2031 data was assumed to represent 2030 (see Table A 1 in the Appendix for full details).

The data point for 2024 is chosen to represent the baseline for thermal demand headroom and future data points are chosen to be for the years 2030, 2040 and 2050. In nearly all cases the DNOs explore the level of headroom available using scenarios developed by the National Energy System Operator (NESO) [20] or have scenarios that can be considered as broadly equivalent. Three of these scenarios were chosen for all DNOs to reflect different levels of ambition in terms of speed of decarbonisation and level of societal change [19] :

- **Falling Short or Steady Progression**: this represents a business-as-usual pathway towards decarbonisation in GB which is not sufficient to meet the net zero emissions target by 2050.
- **Consumer Transformation**: this scenario meets the net zero emissions target by 2050 through substantial individual engagement and behavioural shifts, alongside a high level of electrification.

- **Leading The Way**: this scenario assumes achieving net zero emissions earlier than 2050 due to the widespread use of electric and hydrogen technologies with significant consumer engagement.

## 2.2. Capacity needs for industrial sites

The NZIP model was used by the UK Government to develop its 2021 Industrial Decarbonisation Strategy and by the Climate Change Committee (CCC) in 2020 to inform its sixth carbon budget advice to the UK Government [21]. The model, which is geographically disaggregated, is used in this paper to produce pathways for deep decarbonisation of UK industry, which include results showing future electricity demand from both large industrial sites that are designated as point sources of emissions (individually) and smaller sites that are included in non-point sources of emissions (by region). The increase in electricity generation (in MWh) from 2024 is then converted to an increased need for electrical capacity (in MW) assuming a 90% load factor for the years 2030, 2040 and 2050. For this analysis, only sites from the main manufacturing sectors are considered: Iron and Steel, Chemicals, Cement and lime, Food and Drink, Glass, Paper, Other minerals (such as Ceramics manufacturing), Non-ferrous metals, Vehicles and Other industry.

Three pathways are produced for industrial decarbonisation according to the following scenarios:

- **Balanced:** it is the main pathway considered for the 6$^{th}$ carbon budget analysis by the CCC which implements low-regret actions and decarbonisation technologies in the long-term.
- **No Resource and Energy Efficiency (REEE)**: The NZIP model assumes a significant role for REEE to decarbonise industrial sectors such that sites often use less electricity in the future compared to their baseline even when adopting electrification technologies. As such, this scenario represents a sensitivity analysis for the 'Balanced' pathway but with no REEE measures.
- **Max electrification**: this represents a pathway where the wide implementation of electrification technologies is encouraged with low electricity prices. This is designed to reflect the increased role that electrification could play in decarbonising UK industry if supportive policies are introduced in the future.

Figure 2 shows the additional electric capacity needed by industrial sectors across the different decarbonisation pathways (calculated as the capacity needed in 2050 minus the baseline capacity in 2024). It shows that Chemicals, Food & Drink, and Other Industry sub-sectors dominate the additional electric capacity needs across all three pathways as these sub-sectors are mostly decarbonised with electric technologies (see Figure 3). The additional capacity needed for these three sub-sectors accounts for around 4 GW compared to only 0.8 GW from other sub-sectors in the Balanced scenario. This capacity rises to around 5.4 GW and 6.8 GW for these three sub-sectors compared to 1.5 GW and 1.9 GW for the other sub-sectors in the Max Electrification and No REEE scenarios respectively.

Figure 3 shows the additional electric capacity needed by industry comes mainly from the direct use of electricity, rather than its use to produce hydrogen or to power carbon capture and storage (CCS) and bioenergy with CCS (BECCS). The choice of decarbonisation technology in the NZIP model is dependent on many factors, including the availability of decarbonisation options in each sector, their techno-economic parameters (cost, efficiency etc), and other assumptions including energy costs and carbon prices.

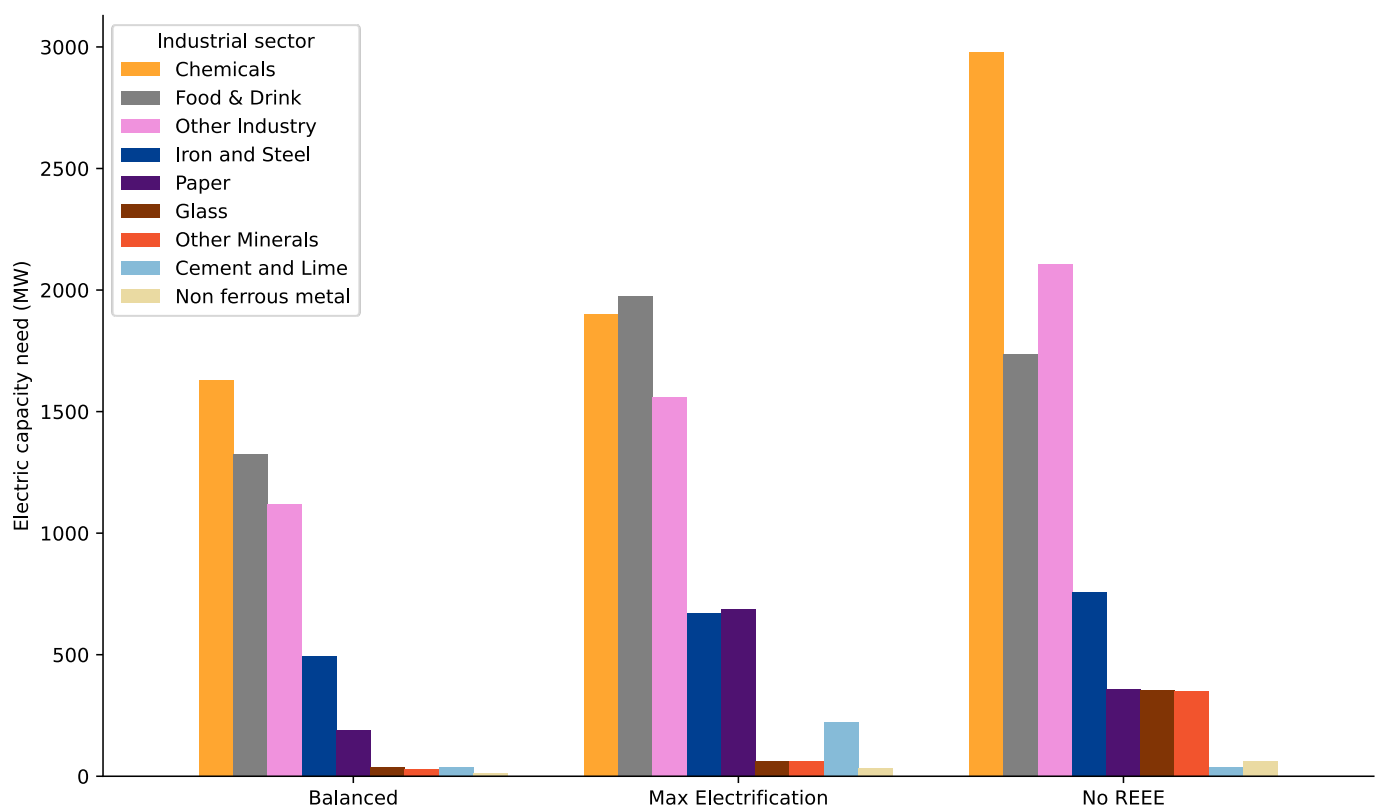

*Figure 2 Additional electricity capacity demand in 2050 for industrial sectors across decarbonisation pathways.*

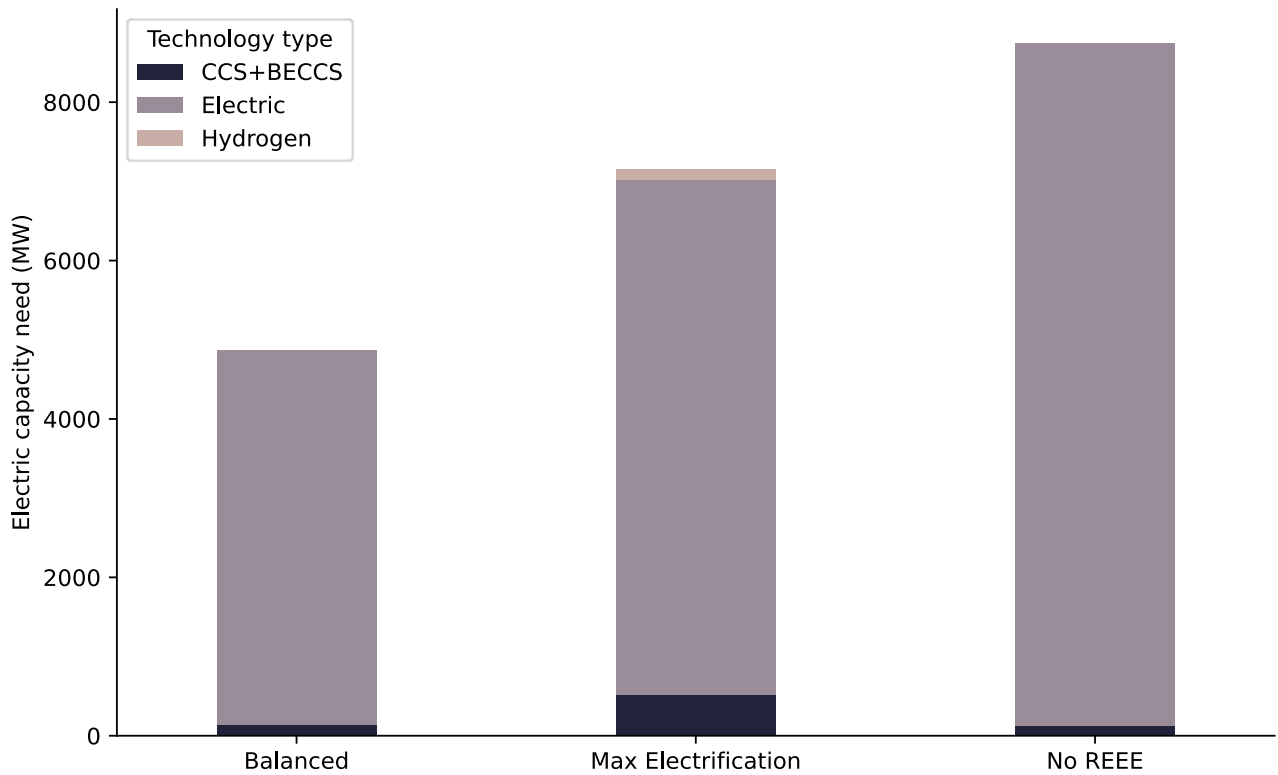

*Figure 3 Additional electricity capacity demand in 2050 for industrial sectors by technology types across decarbonisation pathways.*

## 2.3. Optimisation process for nearest network substation

The location is known for industrial sites that are large point sources of emissions, which means that it is possible to find the nearest substation using the following approach. For each industrial site $S_i$ with geographical coordinates $(Lat_i, Lon_i)$, the nearest substation $N_j$ with coordinates $(Lat_j, Lon_j)$ is found based on the Haversine formula (1) which computes the great-circle distance between two points on the Earth's surface:

$$d_{ij} = 2R \cdot \sin^{-1}\left(\sqrt{\sin^2\left(\frac{\Delta Lat}{2}\right) + \cos(Lat_i) \cdot \cos(Lat_j) \cdot \sin^2\left(\frac{\Delta Lon}{2}\right)}\right) \ldots (1)$$

Where $R$ is the Earth's radius (6371 km), $\Delta Lat = Lat_j - Lat_i$ and $\Delta Lon = Lon_j - Lon_i$ are the differences in latitude and longitude between the site and the substation. This distance is used to find the nearest substation $N_j^*$ for each site as in (2):

$$N_j^* = \arg\min_j \ d_{ij} \ldots (2)$$

For each site connected to a substation, the constrained capacity decision depends on both the remaining capacity of the substation and the site's capacity. The constrained capacity $C_{i,y,s}^{Consntrained}$ in site ($i$), year ($y$) and scenario ($s$) is determined according to the following conditions:

$$C_{i,y,s}^{Consntrained} = \begin{cases} \max(P_{i,y} - C_{j,y,s}^{Headroom}, 0) & if\ C_{j,y,s}^{Headroom} > 0\ and\ P_{i,y} > 0 \\ |C_{j,y,s}^{Headroom}| + P_{i,y} & if\ C_{j,y,s}^{Headroom} < 0\ and\ P_{i,y} > 0 \\ 0 & Otherwise \end{cases}$$

Where $C_{j,y,s}^{Headroom}$ is the available headroom capacity of substation ($j$) in year ($y$) and scenario ($s$) and $P_{i,y}$ is the power demand of site $S_i$.

After determining the constrained capacities, an optimal allocation strategy for sites that are connected to the same substation $S_j$ is implemented such that sites with the smallest capacity needs are allocated first, and the substation's remaining capacity is reduced accordingly:

$$C_{j,y,s}^{Headroom} = C_{j,y,s}^{Headroom} - P_{i,y} \quad for\ S_i \in S_j$$

This approach minimises the number of sites that are found to be constrained, and so can be thought of as a "best-case" scenario.

## 3. Results

### 3.1. Electric capacity needs for all industrial sites and implications for network headroom

When considering the 'Balanced' scenario, Figure 4 shows that the available network headroom in 2030 is sufficient to meet the extra capacity demands from industrial sites in that year in nearly all GB regions and network scenarios. After 2030, the expected decarbonisation pace accelerates resulting in higher electrical capacity needs from industrial sites for 2040 and 2050 compared to 2030. Without further investment beyond 2030, the network would become significantly constrained in the central, south, and north-west areas of England, and Wales. This is mainly due to decreased headroom in the DNO scenarios from the predicted increase in electricity demand from non-industrial sources (e.g. heat pumps and electric vehicles), rather than because of the increase in industry electricity demand. If the 'Leading The Way' scenario is considered for 2040 and 2050, then the total network headroom across the whole of GB is significantly negative (i.e. constrained), with a shortfall of capacity of between 44 GW and 63 GW respectively. The contribution of industry to the decreased headroom is quite modest: the additional capacity needs from industrial sites over the same period across constrained regions is around 4 GW. Nevertheless, as we show later, the implications for industrial decarbonisation are significant.

The level of future network headroom varies depending on the underlying assumptions for electricity demand growth predicted by the DNOs. By 2030, the network headroom is around + 44 GW, +39 GW, and +29 GW in the 'Falling Short', 'Consumer Transformation', and 'Leading The Way' scenarios respectively. This headroom decreases significantly by 2050, becoming negative at the GB level under all scenarios with capacity shortfall of 24 GW, 71 GW, and 63 GW for the three scenarios respectively. Even in regions with sufficient network headroom in 2050, such as London and the north-east of England, the increase in industrial capacity needs (while relatively modest) often represents a significant proportion (between 15-50%) of the total available network headroom indicating that even a slight additional increase in overall electricity demand (whether from industry or other sectors) could also lead to these regions being constrained.

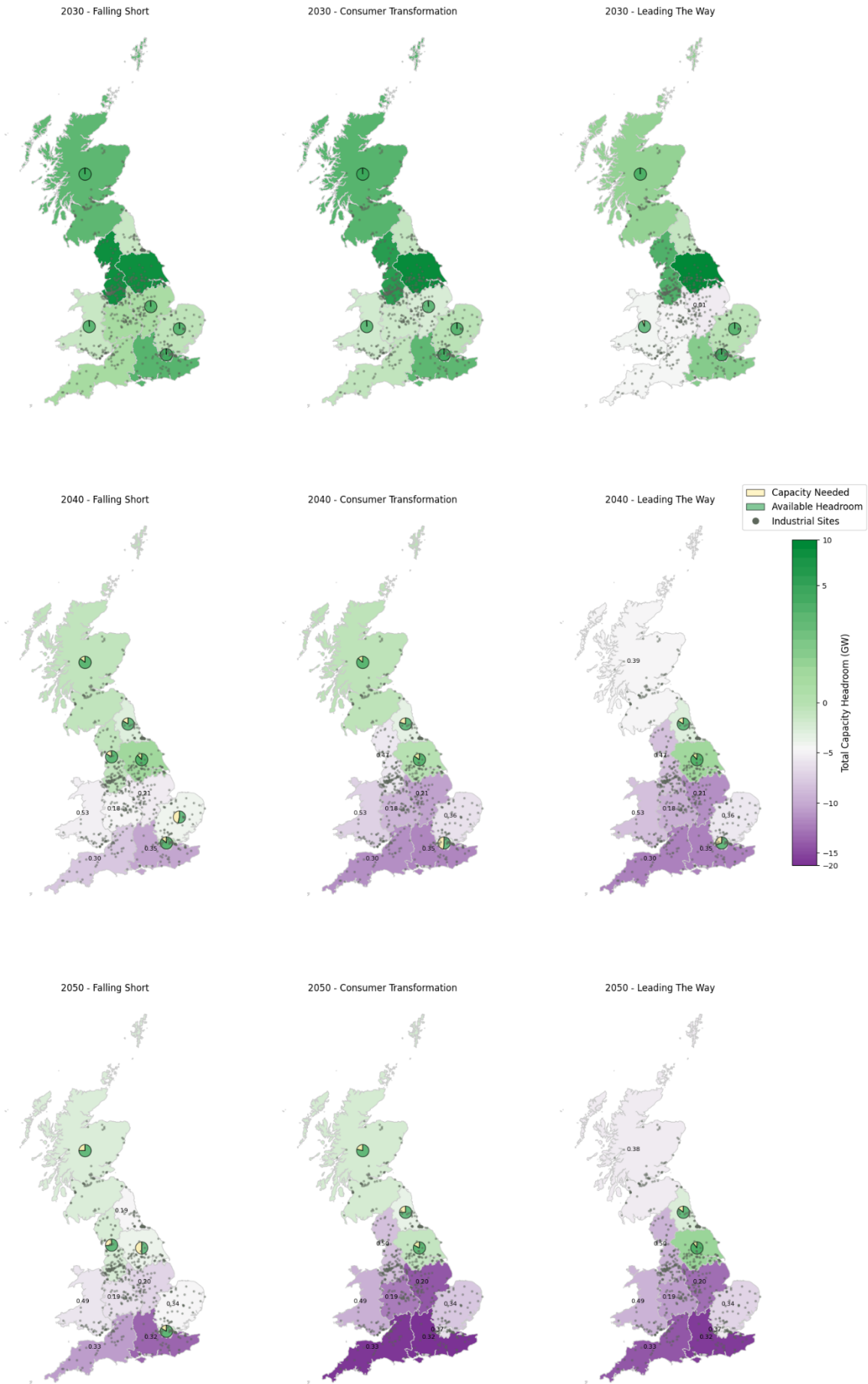

*Figure 4 Capacity needs for industrial sites and network headroom across GB – Balanced scenario. Pie-charts represent the capacity needed by industrial sites (yellow) compared to the available headroom in a region. The numbers inside the regions represent the extra capacity needed by industrial sites in GW when no headroom capacity is available.*

To understand the significance of these potential capacity constraints for emissions we use the Balanced scenario to calculate the level of industrial emissions in 2030 that are associated with industrial sites that are constrained in 2050. The reason for choosing 2030 is that, in the short-term, most sites are not constrained and so emissions in 2030 are reduced from their baseline (2024) levels through a combination of REEE and electrification. This approach therefore gives a fairer estimate of the emissions reductions that are at risk if the network constraints are not addressed than using 2024 data. The results show that the remaining emissions in 2030 from the sites that are constrained in 2050 are between 24 and 29 $MtCO_2e$ (Figure 6), which is nearly 50 to 60% of total industrial emissions in 2030.

When considering the 'No REEE' and 'Max electrification' scenarios (see Figures A.2 and A.3 in the Appendix), the expected level of capacity needed by industrial sites rises across GB by 60% and 123% respectively in 2050 compared to the 'Balanced' scenario. This reduces the available headroom and increases the potential network constraints particularly in the Midlands and Wales. Figure 5 presents the level of new network capacity that would be needed to address these constraints (calculated as DNO network headroom – capacity needs by industrial sites) by 2050 across different network and decarbonisation scenarios. It shows that the level of additional grid capacity required increases from around 30 GW in 'Falling Short' scenario to more than 70 GW in 'Leading The Way' while reaching a maximum of 80 GW when 'Consumer Transformation' and 'Max Electrification' are considered. It should be noted that these levels of additional grid capacity are as a result of increases in demand from all sectors, not just industry, as solving the grid constraints would benefit all electricity users.

Without these increases in network capacity, the industrial emissions in 2030 that are associated with constrained sites in 2050 range between 24 to 40 $MtCO_2e$ as shown in Figure 6 compared to total emissions from industry of 48.5 $MtCO_2e$ in 2030 with higher expected residual emissions in the 'No REEE' and 'Max electrification' scenarios compared to the balanced scenario. This represents nearly 50-82% of the total expected industrial emissions by 2030.

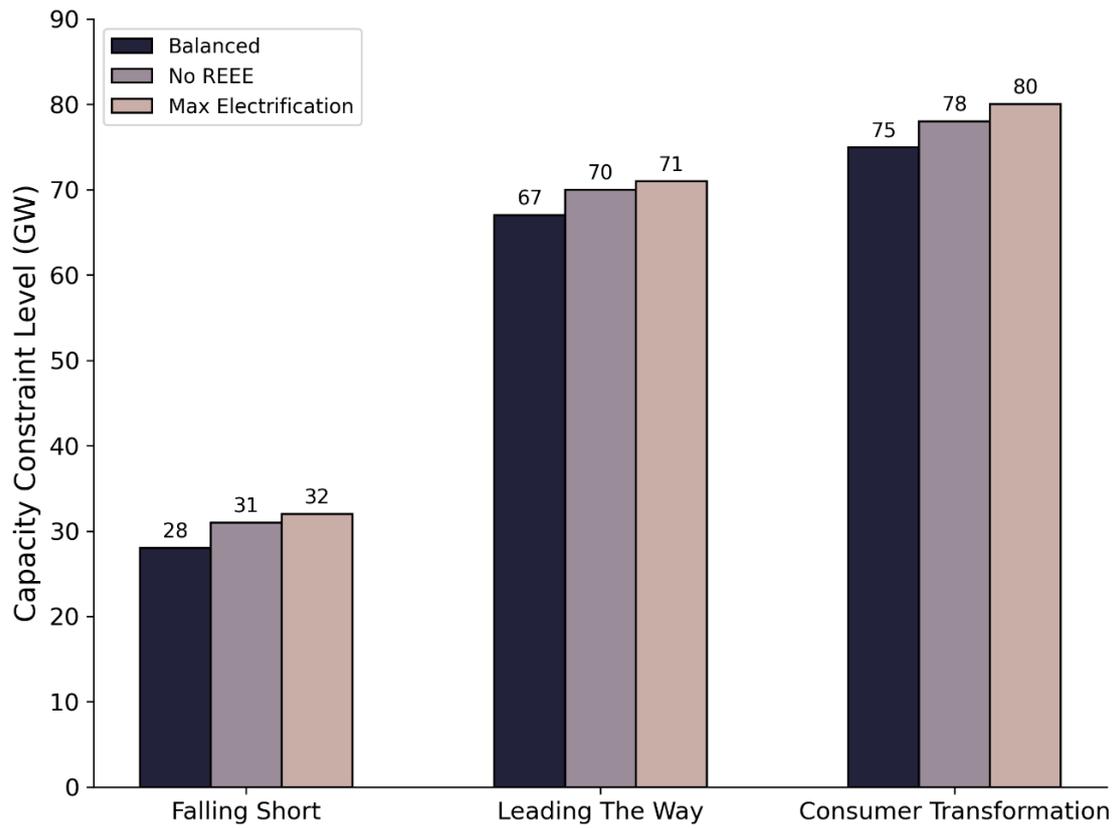

*Figure 5 Estimated electric capacity constraint level in GB by 2050.*

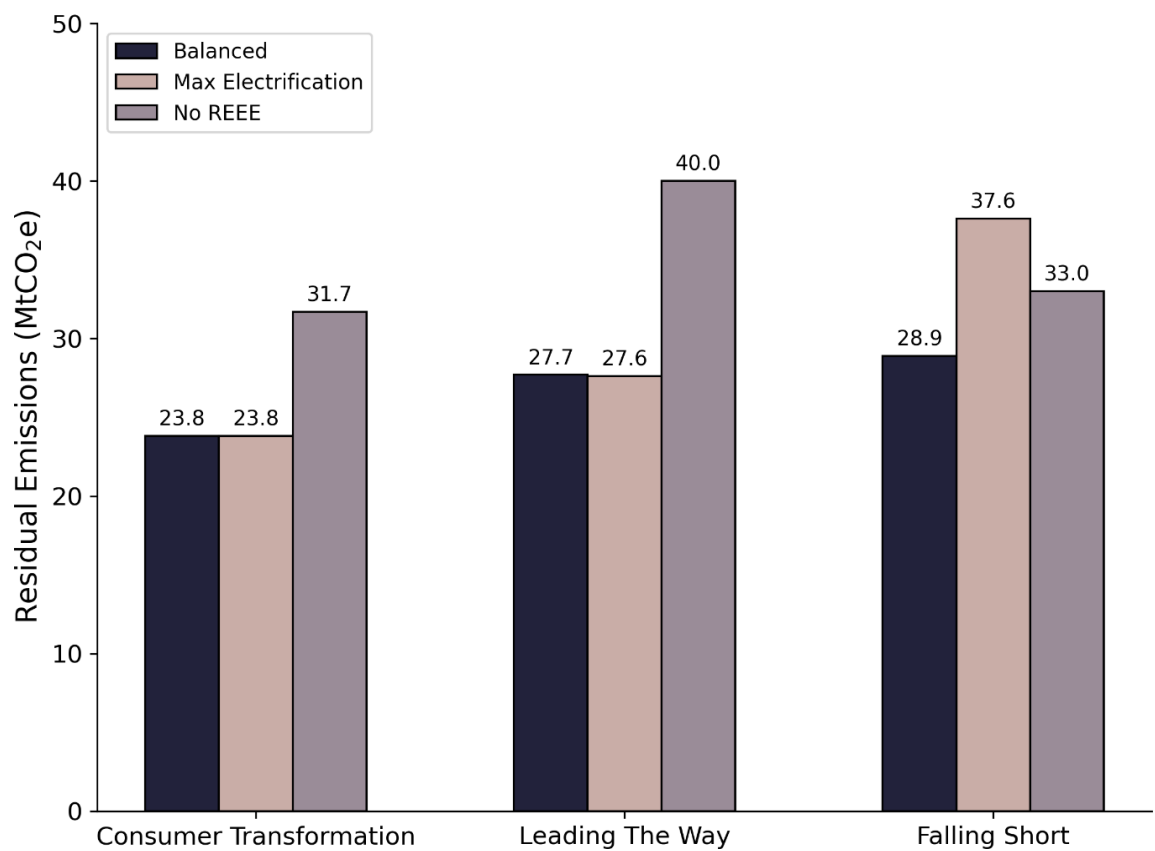

*Figure 6 Estimated residual emissions in MtCO$_2$e by 2030 considering regional constrained electric capacity levels.*

## 3.2. Electricity capacity needs for large (point source) industrial sites only

This section presents key results for large (point source) industrial sites considering the 'Balanced' scenario only, while sensitivity analyses are presented in section 3.4 using the other two scenarios (No REEE and Max Electrification).

### 3.2.1. Percentage of industrial sites that are constrained and the total capacity needed

Figure 7 shows that, without additional network investment after 2030, the percentage of large industrial sites that would face electricity constraints rises significantly from around 20% by 2030 to reach around 65% (425 sites out of a total of 654) by 2040 and 2050 across the different network scenarios as industrial decarbonisation accelerates according to the 'Balanced' pathway. Around 75% of these sites would be constrained as the nearest substation is already constrained due to rising demand from other sources, while the remaining sites are constrained due to a lack of sufficient headroom at the nearest substation to meet the additional electricity demand from the site. Around 30% of sites tend to have the 'same' nearest substation due to the location proximity of sites to each other in certain regions which reduces the available network headroom capacity.

Figure 8 shows that the additional network capacity needed to alleviate constraints rises from a peak of 1 GW by 2030 to around 10 GW by 2040 and 13 GW by 2050 considering the 'Consumer Transformation' and 'Leading The Way' scenarios respectively. In comparison, only half of this capacity is required across the years if the 'Falling Short' scenario is considered, as it reflects a slower pace of electrification (i.e., business-as-usual) and greater continued reliance on fossil fuels. The additional capacity needed to address the electricity constraints on large sites by 2050 is therefore significantly lower than the 80 GW figure found for both large (point) and smaller (non-point) sites when these are considered at a regional level (see Figure 4).

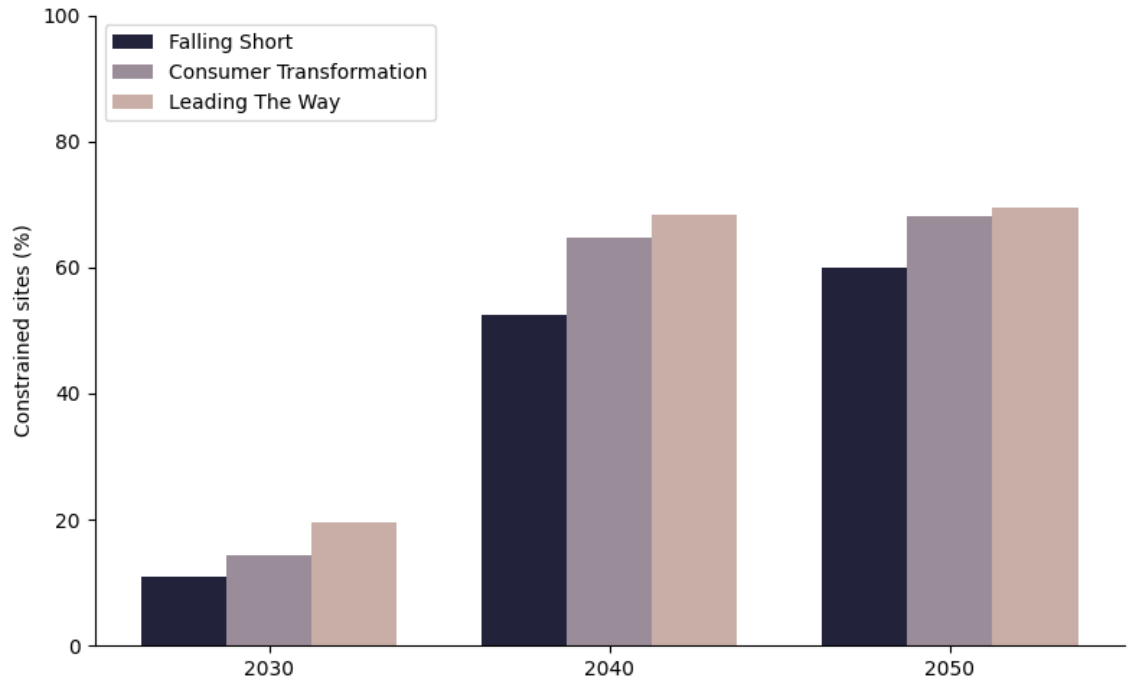

*Figure 7 Percentage of industrial point-sources constrained sites with total sites = 654.*

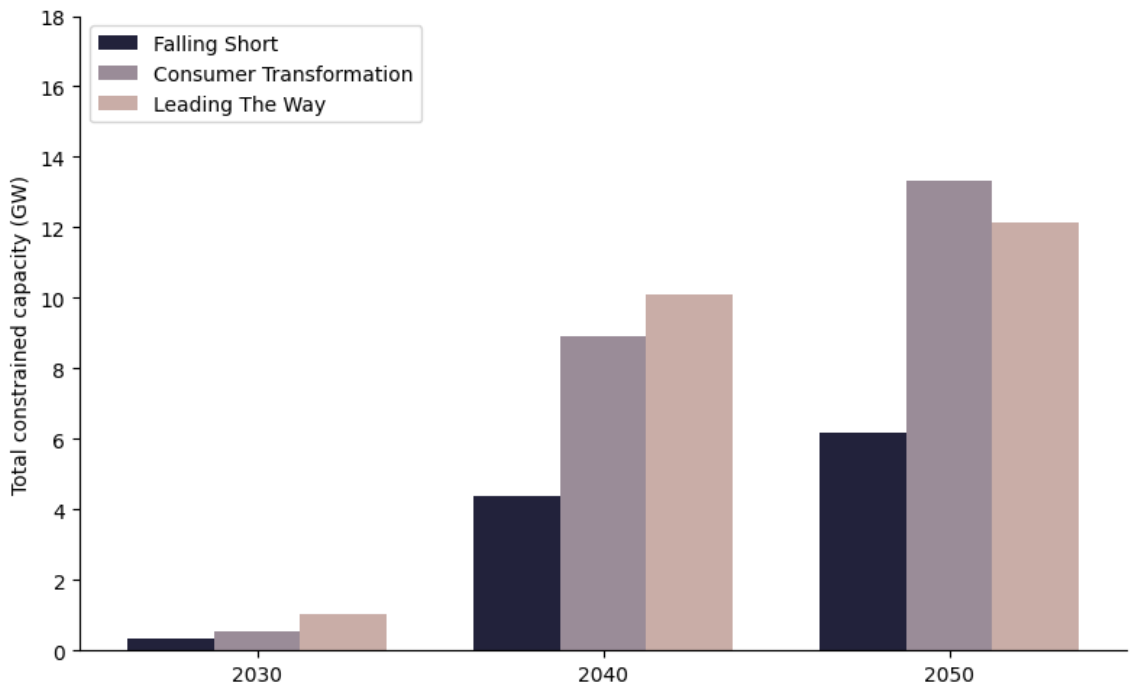

*Figure 8 Total capacity needs for constrained point-sources industrial sites.*

## 3.2.2. The additional electric capacity needs for constrained large industrial sites

Figure 9 shows that large industrial sites are mostly constrained because the nearest substation is already constrained (due to predicted demand increase from other sectors) accounting for 319, 385, and 413 sites across Falling Short, Consumer Transformation, and Leading The Way respectively. In comparison, around 73, 61, and 41 sites are constrained as there is not enough headroom in the nearest substations across Falling Short, Consumer Transformation, and Leading The Way scenarios respectively. For both constraint reasons, the additional electric capacity needed from constrained large industrial sites (i.e., without accounting for the capacity needed to upgrade the substations that are already constrained) can reach around 2 GW for all three network scenarios. It should not be confused with the results of Figure 8 which represents the extra capacity needed at constrained substations to meet the increased demand from all sectors, not just large industrial sites.

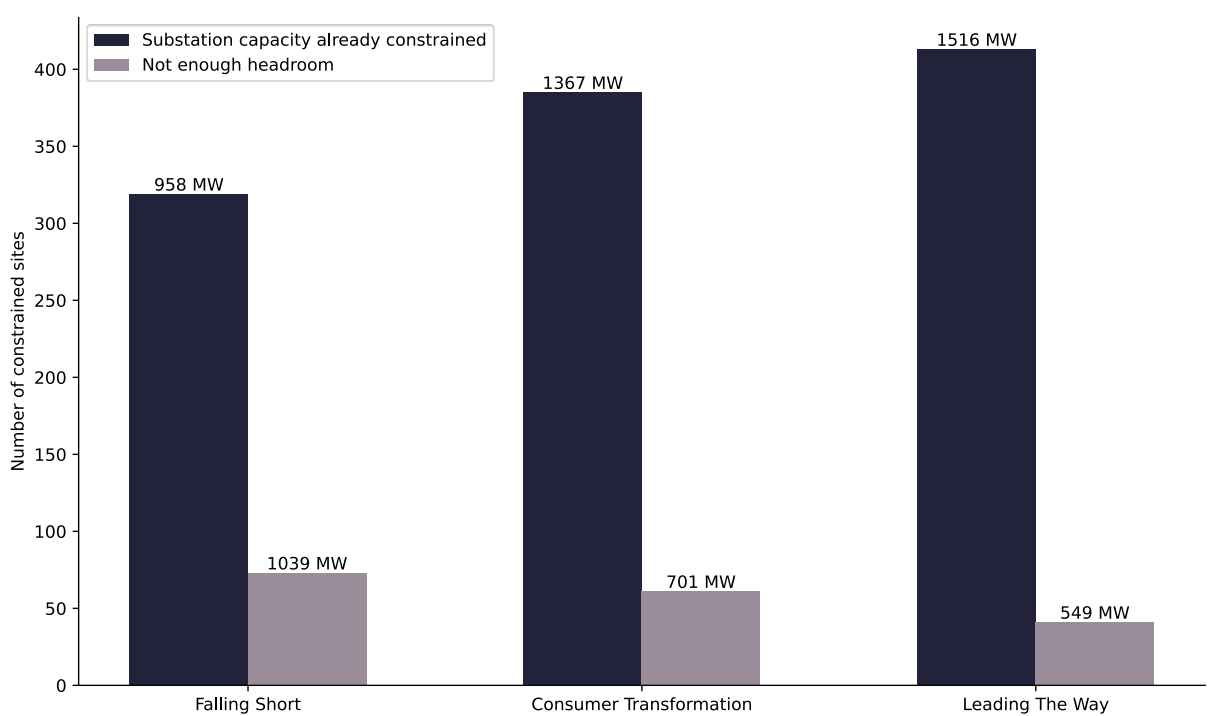

*Figure 9 Number of constrained sites by reason and industrial capacity needs in 2050.*

### 3.2.3. The impacts of constraints on large industrial sites' y emissions and location

Figure 10 shows the percentage of 2030 emissions for each sub-sector of industry that are associated with large industrial sites (point sources only) that could be constrained in 2050. It is calculated as the 2030 emissions for potentially constrained large sites in a sub-sector divided by the total emissions for both constrained and non-constrained sites in the same year (e.g. total sub-sector emissions for point sources).  It reveals that sites that belong to the Iron and Steel, Food & Drink, Chemicals, Glass, and Other Minerals sub-sectors could be most affected, reflecting a combination of the lack of network headroom at the nearest substation and the increased demand for electricity from the decarbonisation options that are implemented by the site. Overall, the total 2030 emissions associated with potentially constrained sites in 2050 across all sectors in Figure 10 could reach around 24 MtCO$_2$e compared to emissions total of 35 MtCO$_2$e from all large industrial sites (i.e., 69 % of large industrial point sources sites) across all network scenarios.

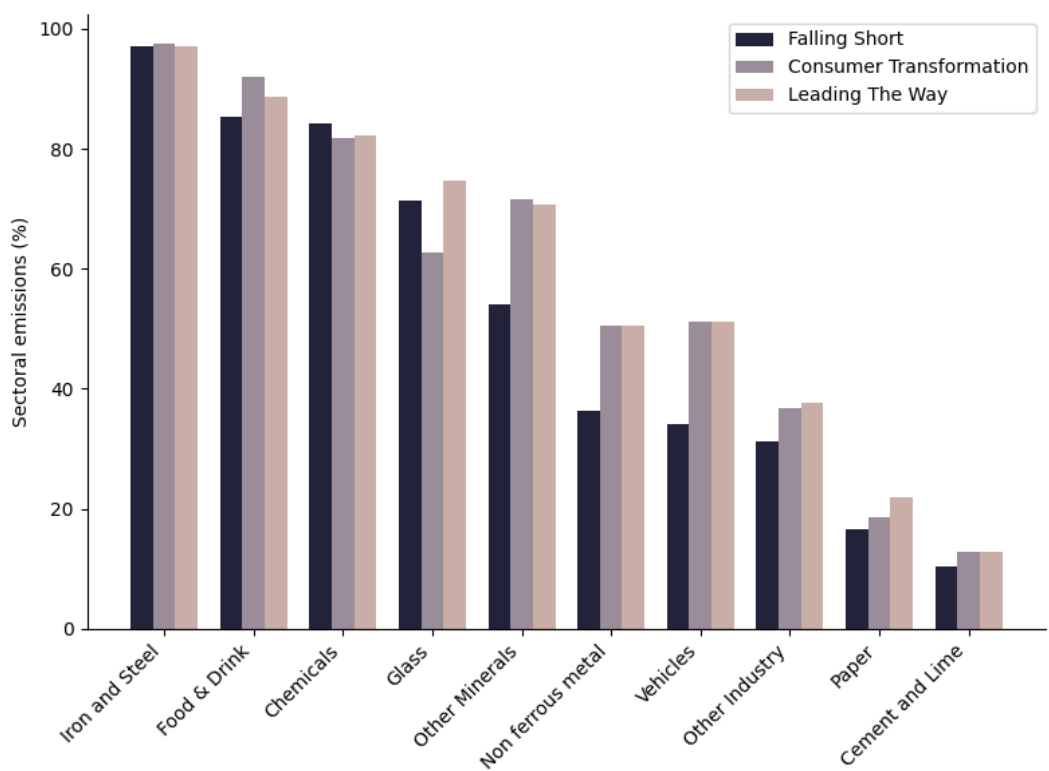

*Figure 10 Industrial sector baseline emissions in 2030 for constrained point-sources sites in 2050.*

The location of potentially constrained industrial sites by sub-sector in 2050 (for the Leading The Way scenario) is shown in Figure 11. The majority of these sites are located in England and Wales, with fewer in Scotland. In general, the location of the sites is aligned with the regions that could have the least available network headroom in 2050 without further network strengthening (see Figure 3). Food & Drink, Other Minerals, and Chemicals sites represent the majority of constrained large industrial facilities with 124,114, and 96 sites respectively.

In UK policy, a distinction is often made between industrial sites that are part of 'industrial clusters'; traditional heartlands of UK industry in which energy intensive industries from different sub-sectors are co-located, often near the coast, and 'dispersed sites', which are industrial sites located at least 25-30 km away from an industrial cluster (see [22] for more details). Using this definition, Figure 12 shows that there are around three times as many dispersed sites that could be constrained compared to cluster sites across the three scenarios. The reasons for this are two-fold; firstly, there are many more dispersed sites (489) than clustered sites (190) in GB and secondly, the available headroom on the network tends to be greater in areas where industrial clusters are located (as shown in Figure 3).

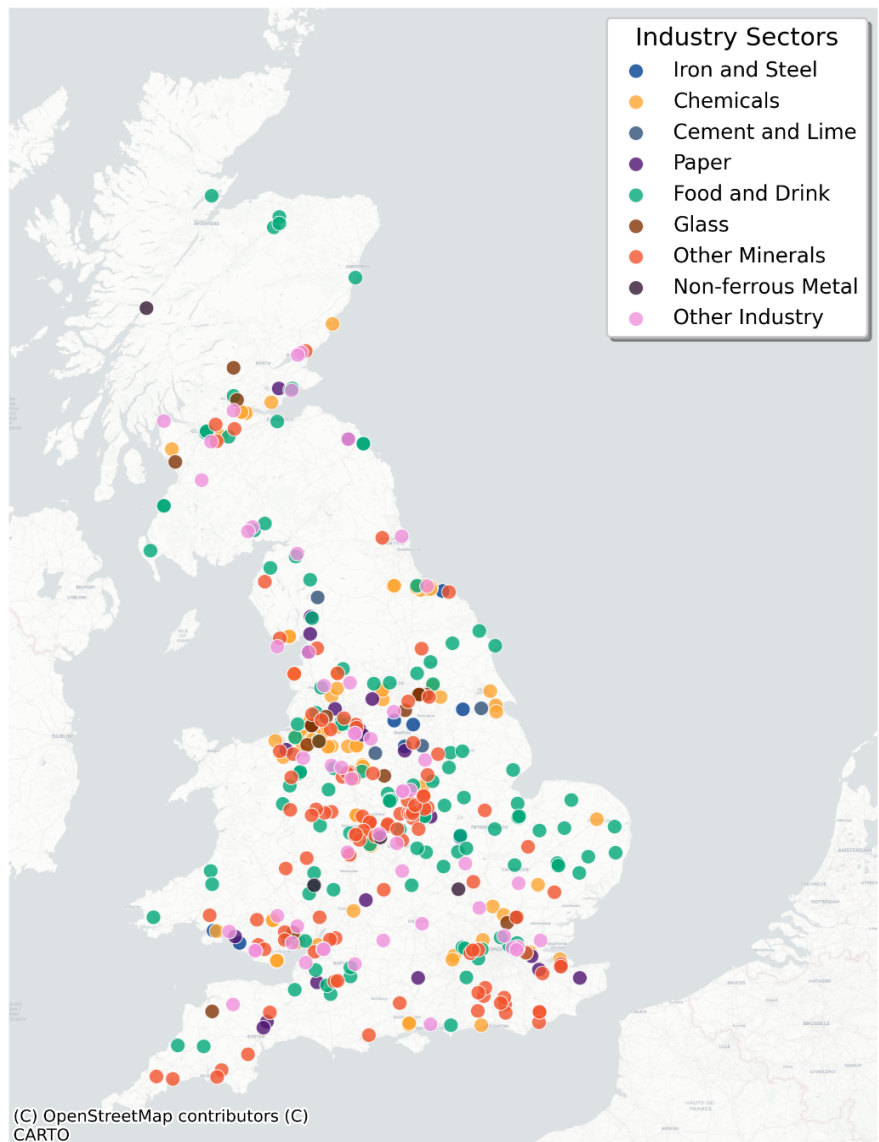

*Figure 11 Location of constrained point source industrial sites by 2050 (Leading The Way) by sector.*

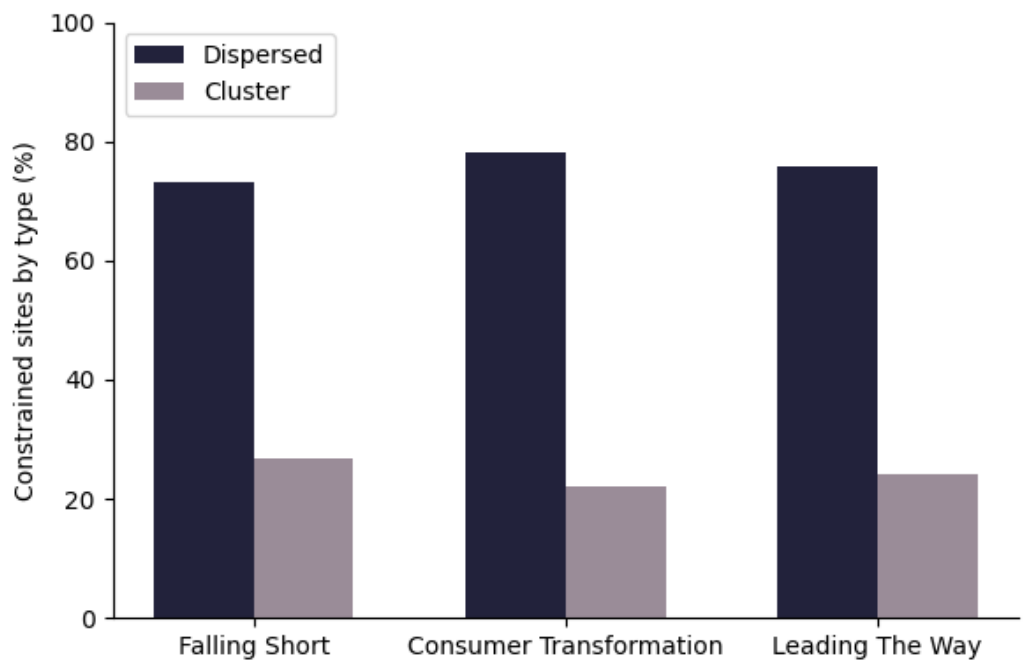

*Figure 12 Location type of constrained point-source industrial sites by 2050.*

## 3.3. Summary of the total capacity needed to enable industrial decarbonisation

Figure 13 summarises the total additional electric capacity needed by industry in GB in 2050 for the Balanced decarbonisation scenario combined with all three electricity network scenarios. The electricity demand for all industrial sectors in 2050 is around 4.8 GW higher than in 2024. This would require an extra 4 GW of distribution network capacity since most of GB's regions are constrained in 2050 (see Figure 4) and therefore only 0.8 GW of this additional demand can be met by the existing network. On the network side, the DNOs network development plans show that the shortfall in capacity for GB's distribution network reaches 24 to 71 GW (depending on the network scenario) by 2050. The total network capacity needs are therefore between 28 and 75 GW to mitigate both these expected network shortfalls (24 – 71 GW) and to meet the additional electricity demand for all constrained industrial sites (4 GW). If only industrial sites that are large point sources of emissions are considered, then the total extra capacity needed for constrained sites is around 2 GW while the shortfall in the electricity capacity in all nearest substations is between 4 and 9 GW, bringing the total additional capacity needed to between 6 and 13 GW. Under more ambitious decarbonisation pathways such as 'No REEE' and 'Max Electrification', the corresponding capacity needs for industrial sites will increase (see section 2.2, 3.2.1, and 3.2.3).

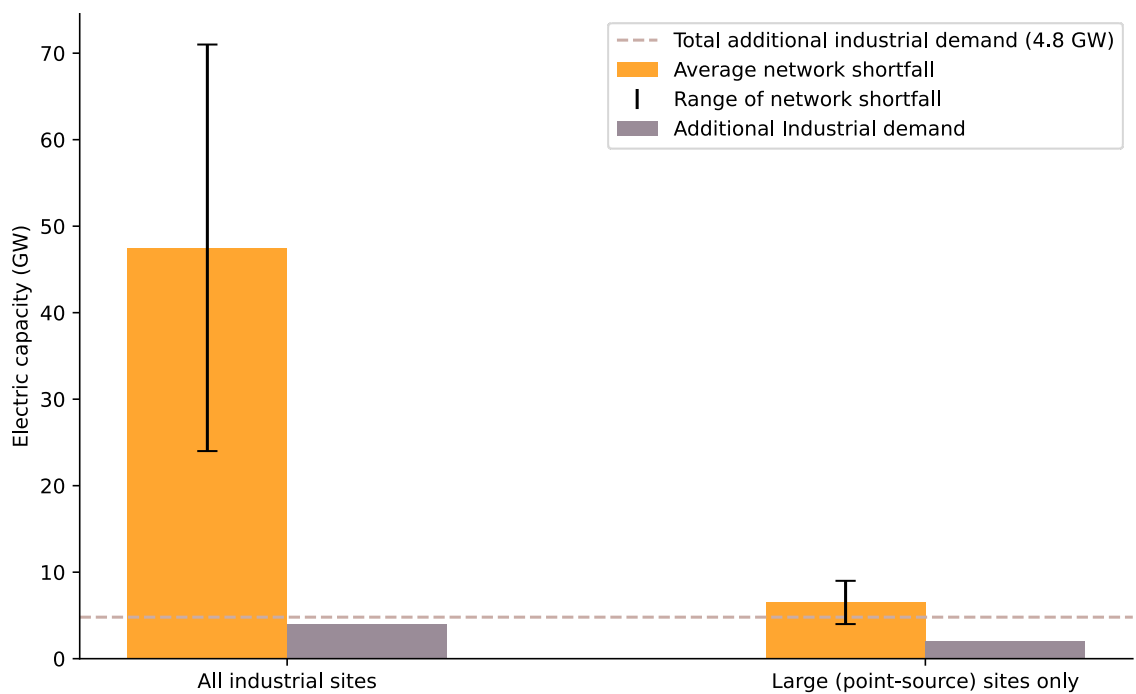

Figure 13 Summary of the total electric capacity needed to enable industrial decarbonisation by 2050 for the Balanced pathway and the three network scenarios.

### 3.4. Sensitivity analysis

Sensitivity analyses were performed to explore how the results in section 3.2 for the 'Balanced' scenario change under the assumptions for the 'No REEE' and 'Max electrification' decarbonisation scenarios. The results shown in Figure 14 are for 'Leading The Way' network scenario for 2050 (results for other network scenarios are in Figure A 4 and A5 in the Appendix). The percentage of potentially constrained sites increases by 6-8% under both alternative scenarios compared to the 'Balanced scenario. This results in an additional capacity requirement of 2-3 GW above that seen in the Balanced scenario to mitigate the network constraints and enable the electrification of industrial sites, bringing the total additional capacity needed to nearly 15 GW by 2050 (see Figure 8 for comparison).

This also means that under the 'No REEE' and 'Max electrification' scenarios, the total 2024 emissions associated with the constrained sites increases by around 7 MtCO$_2$e and 3 MtCO$_2$e respectively by 2050 compared to the 'Balanced scenario'. These results illustrate that more ambitious policy to promote electrification and lower emissions savings from the application of energy and resource efficiency (compared to the Balanced scenario) would both increase the pressure for new network investment and upgrades.

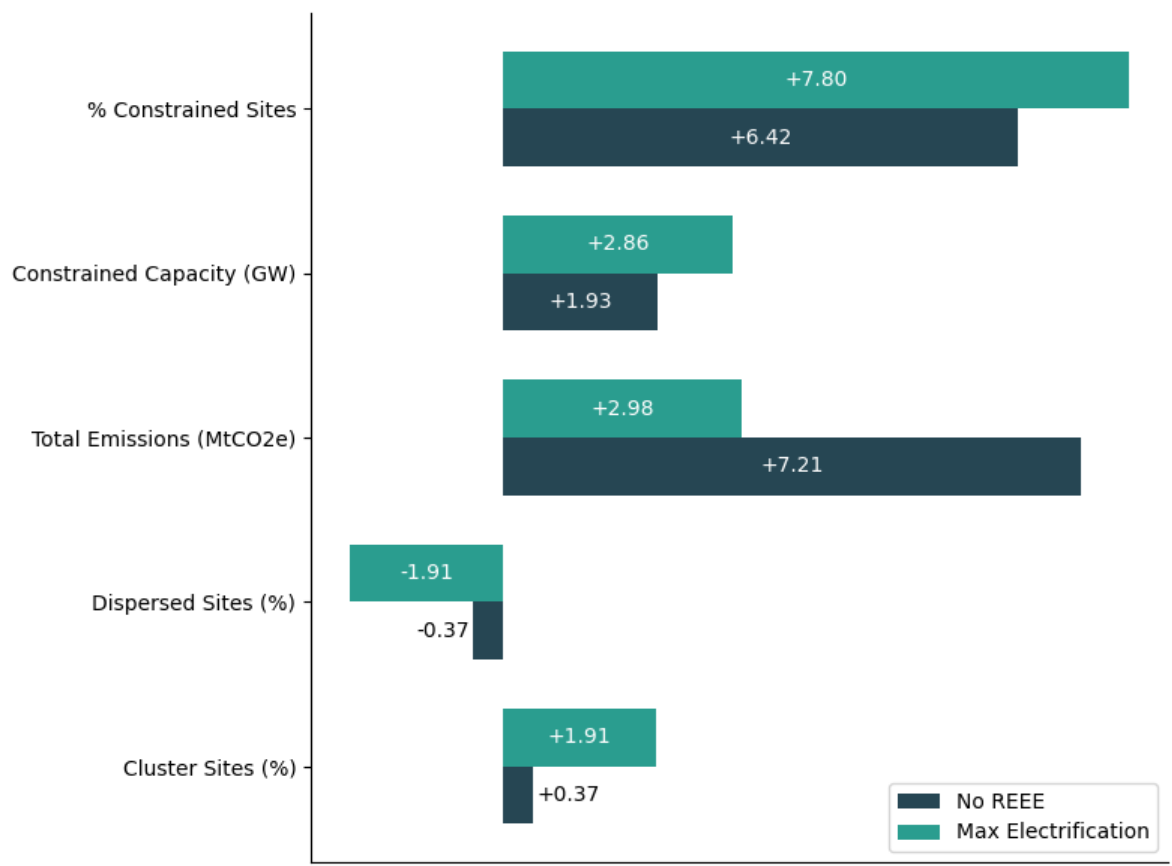

*Figure 14 Sensitivity analysis results for No REEE and Max electrification scenarios and Leading The Way by 2050. Note that the equivalent data for the Balanced scenario results can be found in the following figures: '% Constrained Sites' (figure 7), 'Constrained Capacity (GW)' (figure 8), 'Total Emissions (MtCO2e (figure 10)), and 'Dispersed Sites (%) and Cluster Sites (%) (both figure 12)*

## 4. Discussion

The results show that timely investment in the GB electricity distribution networks beyond 2030 will be essential to accommodate the predicted electricity demand increase across all sectors including manufacturing industry. Future electricity demand could reach 2-3 times the current levels by 2050 according to the CCC [23]. The NESO in GB has calculated that cumulative direct investment of up to £60 billion in just the electricity transmission network will be needed to meet the increase in generation and demand, and to decarbonise the power system by 2030 [24, 25]. However, the total investment level could be even higher if industry implements significant electrification as a means to decarbonise and it does not take account of the extra investment need in the distribution network which is the subject of this paper. Since network operators are closely regulated in GB by the Office of Gas and Electricity Markets, the investment needed to unlock grid constraints will need to be reflected in the relevant price controls.

Since the network is mostly constrained in the central, south, and north-west of England, and Wales after 2030, and three times the number of dispersed sites are likely to be constrained compared to cluster sites, taking a place-based approach to mitigating this constraint is necessary. The UK Government announced a Strategic Spatial Energy Plan to drive investment in energy infrastructure including solving grid connection issues [26]. This should consider the heterogenous processes in manufacturing industry and their energy requirements [5] particularly in Food & Drink, Other Minerals, and Chemicals sectors which have the majority of constrained point-source sites. However, if particular industries are unable to decarbonise due to the long delay in getting an electricity capacity connection, they might relocate to regions with ample network capacity and these regional disparities may present just transition issues [22].

Rather than requesting a larger capacity connection to the grid, some industrial sites could choose to invest in on-site generation to meet their increased demand for electricity. However, in most cases this would also require a network connection to provide additional power at times of peak demand, deliver ancillary services to the network and/or participate in the wholesale electricity market. Furthermore, the time to get a generation connection might be no quicker than for a demand connection. GB's NESO predict significant delays with generation connection with over 59 GW of capacity already in the connection queue (although not all projects in the queue get to the operational stage) with significant numbers of renewable energy plants aiming to connect to the distribution network [25].

The results also highlight that models used to inform policy-making should explicitly consider all the infrastructure needs associated with the deployment of decarbonisation technologies, as these can have important impacts on the most appropriate mix of options to decarbonise a particular sector. The NZIP model used previously by both the Government and the CCC includes hydrogen and carbon capture and storage availability and infrastructure needs in a geographically disaggregated way, but does not use the same spatial approach to consider the network constraints or costs associated with upgrading the electricity network [13].

Our analysis represents the first published study to quantify the potential impacts of distribution grid constraints on industrial decarbonisation. However, it is subject to several limitations. First, the analysis does not consider whether the geographical distribution of industry across GB may change in the future. While many industrial sites have been in the same location for decades, it is possible that future economic trends and/or policy decisions could lead to industry moving location, including abroad. Equally, a successful industrial strategy could lead to new companies locating in GB. Any redistribution of industry could help alleviate the constraints identified or make them worse depending on the locations involved. Secondly, this study did not consider the dynamics of the power network which could impact on exactly which sub-stations are constrained and by how much. It also did not consider the extent to which demand response could have a role in mitigating network constraints, if electricity demand could be made more flexible by adjusting industrial production profiles or if any increase in demand might be met from increased on-site generation. Lastly, our optimisation approach did not consider the available headroom in the second or third-nearest substations to a large industrial site nor could it apply this optimisation approach to smaller industrial sites (non-point sources) since their precise location is not available. As such, these remain areas for future research work.

## 5. Conclusion and policy implications

Achieving industrial decarbonisation is likely to increase electricity demand through the deployment of technologies that either directly electrify existing production processes or indirectly increase electricity demand through the introduction of other abatement options, such as CCUS, which require electricity to operate the technology. However, the need to upgrade the electricity network to accommodate these demand increases is often given less attention than the new infrastructure requirements for hydrogen and CCUS pipelines. This paper assessed the electricity network capacity requirements for industrial decarbonisation in GB.

In this study, the NZIP model is used to predict the future electrical capacity need as industrial sites decarbonise under net zero pathways. This is then compared to spatially disaggregated electricity demand headroom data for all DNOs in GB under different future network scenarios.

The results show that the network headroom is sufficient to meet extra capacity demands from industrial sites over the period to 2030. However, from 2030, increased electricity demand from other sectors and accelerated industrial decarbonisation mean that by 2040 the network is significantly constrained particularly in the central, south, and north-west of England and Wales. The potential shortfall in the network capacity could reach 71 GW by 2050 under the 'Balanced' scenario with the additional capacity needs from industrial sites over the same period totalling 4 GW.

When only considering industrial sites that are large point-sources of emissions, the results show that, unless there is continued investment in the network, the percentage of constrained sites rises significantly from about 20% in 2030 to reach around 65% by 2050 across the different network scenarios, with three times as many dispersed industrial sites being constrained as sites in industrial clusters. These sites are responsible for nearly 69% of 2030 industrial point source emissions.

Our work has a number of important policy implications. First, UK manufacturing is worth £217 bn to the UK economy and directly employs 2.6 million people [27]. With increasing global competition for investment, potential network constraints can create uncertainty for GB industry. If not addressed, this could result in lost opportunities and investment as companies struggle to compete internationally or decide to relocate to countries with better access to cheap electricity. The current network regulations in GB do not sufficiently encourage DNOs to invest in network capacity ahead of need. This should be addressed by future network price controls and the connection process should also be simplified so that waiting times for new connections are reduced. Second, the impacts of network constraints do not affect all industrial sub-sectors or regions of GB equally. Sectors such as Food and Drink, Chemicals, and Other Industry are most likely to be more in need of additional network capacity to support decarbonisation, and the central, south, and north-west of England and Wales are likely to become constrained before other regions. The UK Government is currently considering what additional policy support is needed to encourage electrification and, in doing so, it should ensure that some sectors and regions do not miss out on the electrification opportunities that enhanced policy support can offer. Third, modelling of industrial decarbonisation to inform government policy should explicitly consider the impacts of electricity network constraints as well as other infrastructure requirements such as pipelines for hydrogen and carbon dioxide. The likely substantial increase in electricity demands from industry should be factored more explicitly into the various spatial energy plans being developed by the GB NESO.

## Acknowledgements

We gratefully acknowledge the support of UK Research and Innovation (UKRI), the Engineering and Physical Sciences Research Council (EPSRC), the Natural Environment Research Council (NERC), and the Economic and Social Research Council (ESRC), for funding this study through the UK Energy Research Centre (UKERC, grant ref.: EP/S029575/1)). Peter Taylor would also like to acknowledge support from the EPSRC funded Supergen Energy Networks Hub (grant ref.: EP/Y016114/1).

# Data availability

All the data used in this paper is available from our GitHub repository.

# Declaration of interests

The authors declare no competing interests.

# Appendix

Figure A 1 shows point source industrial sites (red) with the selected nearest substations (blue) and other (not nearest) substations which validates our approach in Section 3.2.

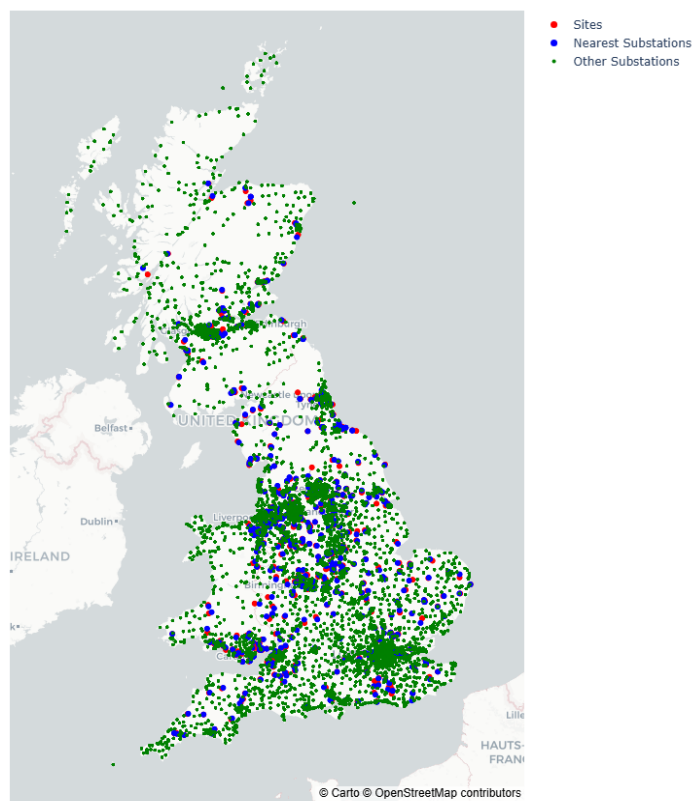

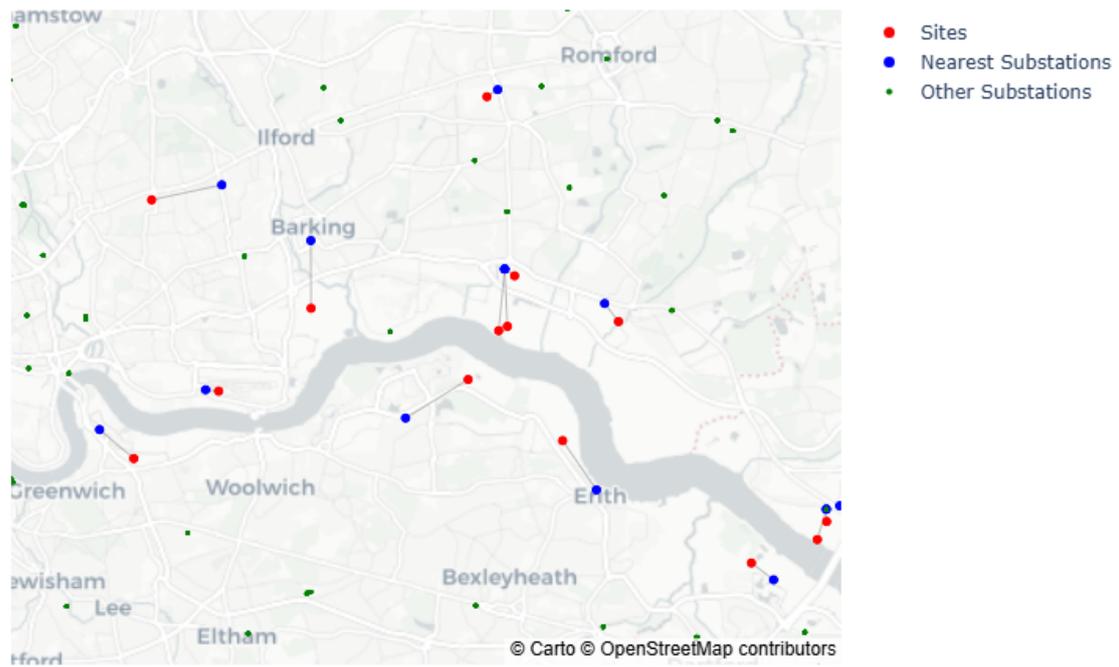

*Figure A 1 Validation of optimisation approach of industrial sites and nearest substation.*

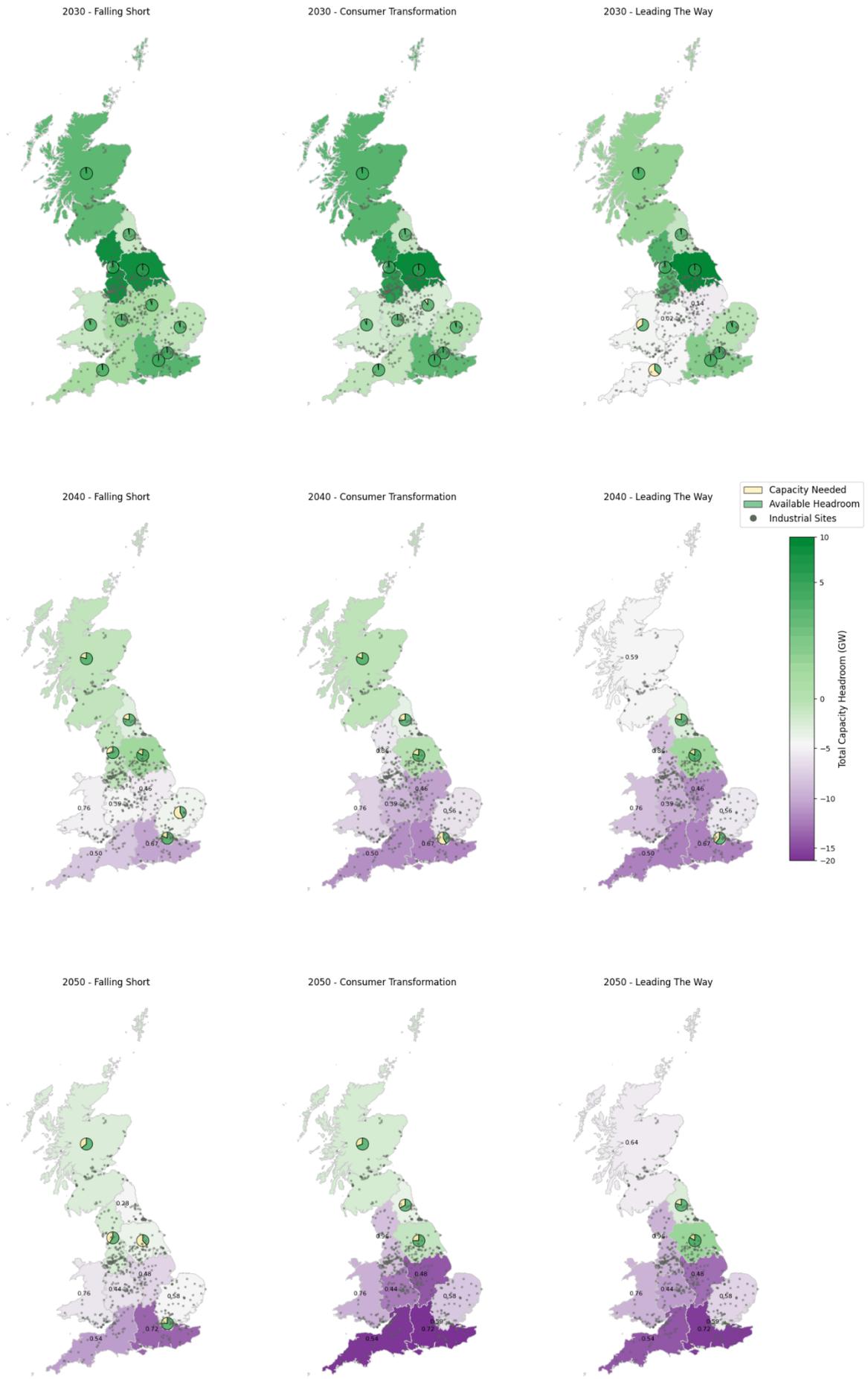

*Figure A 2 Capacity needs for industrial sites and network headroom across GB – No REEE scenario. Pie-charts represent the capacity needed by industrial sites (yellow) compared to the available headroom in a region. The numbers inside the regions represent the extra capacity needed by industrial sites in GW when no headroom capacity is available.*

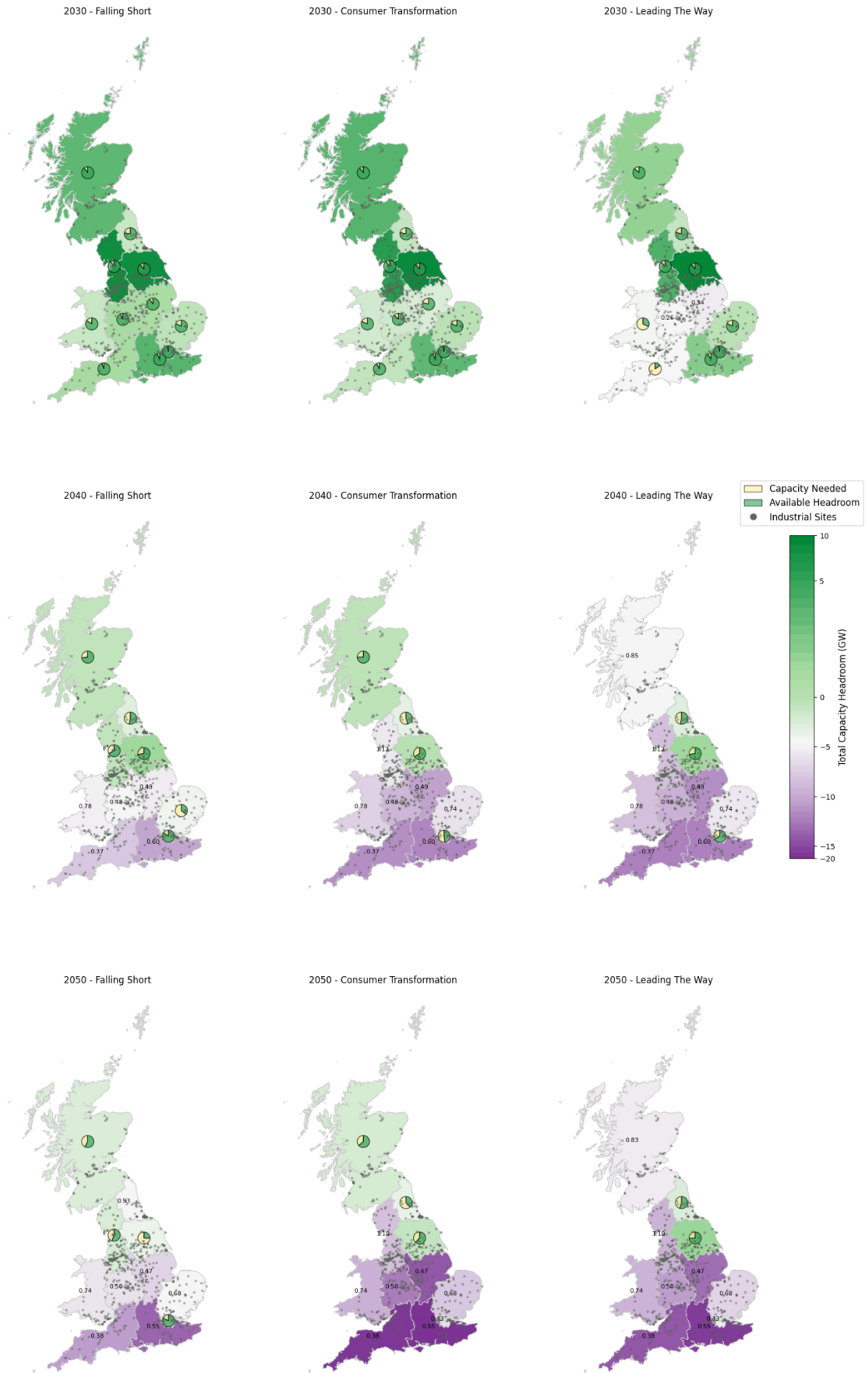

*Figure A 3 Capacity needs for industrial sites and network headroom across GB – Max Electrification scenario. Pie-charts represent the capacity needed by industrial sites (yellow) compared to the available headroom in a region. The numbers inside the regions represent the extra capacity needed by industrial sites in GW when no headroom capacity is available.*

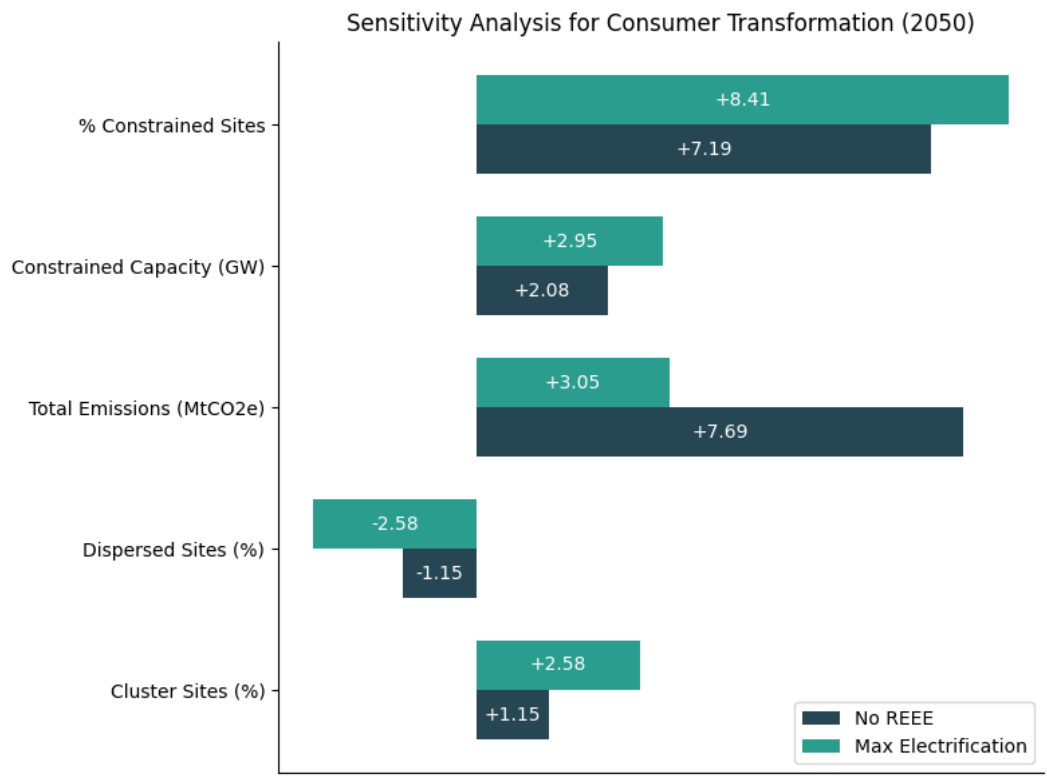

*Figure A 4 Sensitivity analysis results for No REEE and Max electrification scenarios and Consumer Transformation by 2050.*

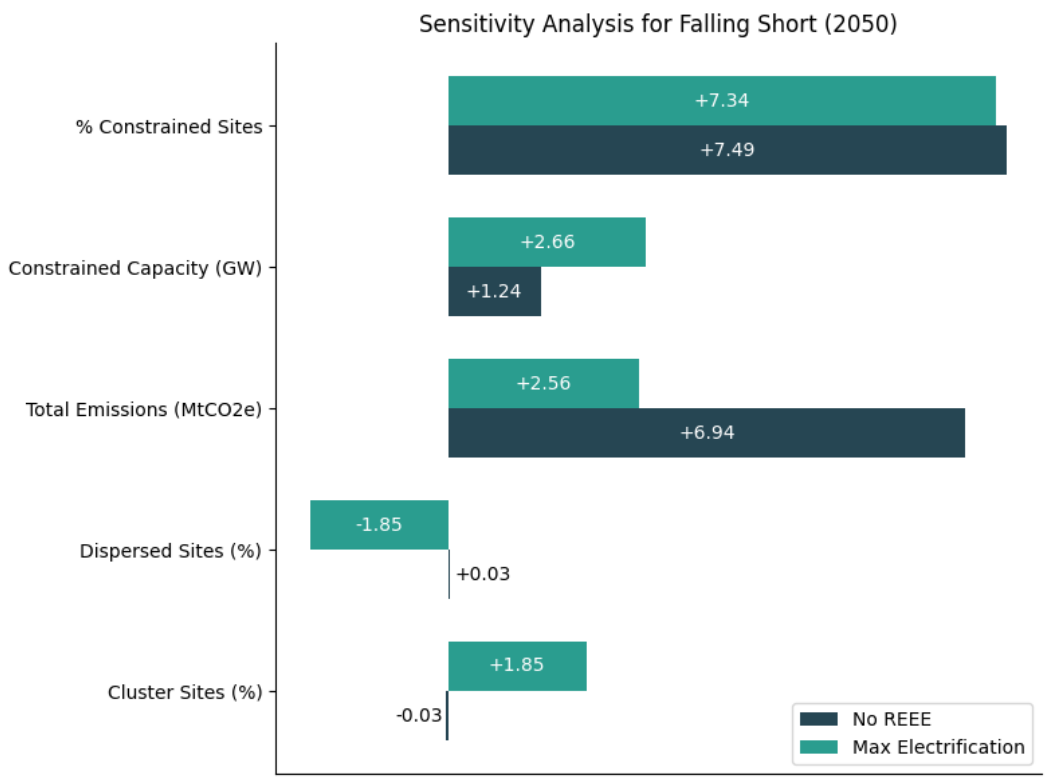

*Figure A 5 Sensitivity analysis results for No REEE and Max electrification scenarios and Falling Short by 2050.*

*Table A 1 Changes made to DNOs data[14-19] in this paper.*

| DNO | Changes made to make the data consistent |
|---|---|
| ENW | - Converted MVA to MW<br>- 2041, 2046, 2051 data was assumed to represent 2040, 2045, 2050 respectively to ensure consistent 'years' data across GB. |
| NPG | - No changes were needed. |
| SPM & SPD | - 'Baseline', 'Low', and 'High' scenarios were assumed to be 'Falling Short', 'Consumer Transformation', and 'Leading The Way' respectively. |
| SEPD & SHEPD | - Converted MVA to MW<br>- Demand Headroom Winter is taken forward rather than Summer and Spring/Autumn. |
| UKPN | - No changes were needed. |
| NGED | - The data comes in 4 separate files for East Midlands, South Wales, South West, and West Midlands and so all these were combined together. |